\documentstyle[12pt]{article}
\newcommand{\beq}{\begin{equation}}
\newcommand{\eeq}{\end{equation}}

\begin{document}
\begin{titlepage}
\begin{center}
{\large {\bf Charged Cosmic String Nucleation in de Sitter Space}}\\
\vspace{1.5cm}

{\bf Minos Axenides}
\footnote{e-mail: axenides@nbivax.nbi.dk}\\
\vspace{0.4cm}
{\em The Niels Bohr Institute\\
University of Copenhagen, 17 Blegdamsvej, 2100 Copenhagen, Denmark}\\
\vspace{1cm}
{\bf Arne L. Larsen}
\footnote{e-mail: allarsen@nbivax.nbi.dk}\\
\vspace{0.4cm}
{\em Nordita\\
17 Blegdamsvej, 2100 Copenhagen \O, Denmark} \\

\vspace{0.4cm}
\end{center}
\vspace{6mm}
\begin{center}
PACS numbers: 0420, 1110, 1117
\end{center}
\end{titlepage}
\newpage
\begin{centerline}
{\bf Abstract}
\end{centerline}
\hspace*{-6mm}We investigate
the quantum nucleation of pairs of charged circular cosmic strings in de
Sitter space. By including self-gravity we
obtain the classical potential energy barrier and
compute the quantum mechanical
tunneling probability in the semiclassical
approximation. We also discuss the classical evolution of charged circular
strings after their nucleation.
\newpage

\section{Introduction}
Inflation \cite{Linde,Rocky,RB} is a
short period of rapid expansion in the early
history of the universe, whereby its presently observable part
originated from a tiny initial region. Topological defects such as
cosmic strings, monopoles and domain walls  are extended objects
present in the spectrum of grand unified theories, that are believed to
be typically generated in phase transitions in the early universe.
They could have acted as seeds for the generation of density
perturbations that resulted in the large scale structure of the universe
and the observed anisotropy in the cosmic microwave background radiation.
Recently it was realized that such objects can be created spontaneously
in a  de Sitter spacetime through the process of quantum
nucleation \cite{Bas}. More specifically, using the static
parametrization it was found that the classical evolution of a circular
string is determined by a simple potential barrier and that strings can
nucleate by a quantum mechanical tunneling through the barrier. In a
realistic situation any topological defects formed at the onset of the
inflationary period are expected to be inflated away. Strings nucleated
towards the inflationary exit will eventually contract upon their
entrance into the radiation dominated phase inside the causal
horizon \cite{JG}.
If they are still circular they are expected to form black
holes \cite{SH}, otherwise they oscillate radiating away their gravitational
energy.

Charged and/or superconducting strings \cite{Wit} have also been the subject of
intense investigation. This is due to their potential role as
seeds of the large scale structure of the Universe \cite{Jer}. Their
dynamical properties have already been analyzed in some
detail (see [9-15] and references therein). Their
electromagnetic structure is
typically represented by a physical degree of freedom in the form of a
world-sheet real scalar field $\Phi.$ In a more recent
analysis of their evolution in de Sitter spacetime, the equations of
motion of their circular loop radius was obtained in the form of an
"effective" repulsive classical potential \cite{JK}. A quantum mechanical
process, where a string of radius $r_1$ tunnels through a local barrier on
the overall repulsive potential and
becomes a string of radius $r_2\neq r_1,$ was then considered and the
tunneling probability was evaluated.

In the present paper we push this picture a little further, in discussing
the question of quantum {\it nucleation} of
(pairs of) charged strings in de Sitter space.
We carefully consider the interplay of the
self-gravitational attraction and coulombic repulsion of small circular
charged string loops. We will argue for the presence of an overall classical
potential energy barrier in the radial equation of motion of the string
loop in de Sitter spacetime.
In a first approximation we subsequently compute the probability
for quantum creation of charged strings and analyze their
classical evolution.

The paper is organized as follows: in Section 2 we derive the charged circular
string evolution equations in de Sitter spacetime. In Section 3 we include the
string self-gravity to a first approximation and calculate the
nucleation probability for charged strings. Finally we consider briefly the
classical evolution of charged strings after nucleation.

\section{String Evolution Equations}
\setcounter{equation}{0}
In this section we present the charged string model and derive the
equations of motion for a uniformly charged circular string in the
background of de Sitter space.

We will consider strings which in the general case are described by the
action:
\begin{equation}
S= -\int d\tau\;d\sigma\;\sqrt{-\mbox{det}G}\;[\mu+\frac{1}{2}
G^{\alpha\beta}(\Phi_{,\alpha}+A_{\mu}X^{\mu}_{,\alpha})\;
(\Phi_{,\beta}+A_{\mu}X^{\mu}_{,\beta})],
\end{equation}
where $\mu$ is the string tension, $G_{\alpha\beta}$ is the induced metric
on the world-sheet, $\Phi$ is a scalar field on the world-sheet and
$A_{\mu}$ is the external electromagnetic potential. The action $(2.1)$ is
well studied in the literature [9-15].
It is however usually expressed in
the "adjoint" form:
\begin{equation}
S^{\dagger}= -\;\int d\tau\;d\sigma\;[\mu \sqrt{-\mbox{det}G} + \frac{1}{2}
\sqrt{-\mbox{det}G}\;G^{\alpha\beta}\Psi_{,\alpha}\Psi_{,\beta}+
\epsilon^{\alpha\beta}A_{\mu}X^{\mu}_{,\alpha}\Psi_{,\beta}],
\end{equation}
where $\Phi$ and $\Psi$ are related by:
\begin{equation}
\epsilon^{\alpha\beta}\Psi_{,\beta}=
\sqrt{-\mbox{det}G}\;G^{\alpha\beta}(\Phi_{,\beta}+A_\mu X^{\mu}_{,\beta}).
\end{equation}
The above model (in the form of eq.(2.1) or (2.2)) should
be considered as a covariant version of the
charged and current carrying superconducting cosmic string discovered by
Witten in a $U(1)\times U(\tilde{1})$ gauge theory \cite{Wit}.

{}From the action $(2.1)$ we obtain the following expression for the
spacetime electromagnetic current:
\begin{equation}
J^{\mu}= -\sqrt{-\mbox{det}G}\;G^{\alpha\beta}\;
(\Phi_{,\alpha}+A_{\nu}X^{\nu}_{,\alpha})\;
X^{\mu}_{,\beta}
\equiv j
\frac{\partial X^\mu}{\partial\sigma}-\rho\frac{\partial X^\mu}{\partial\tau},
\end{equation}
where j is given by:
\begin{equation}
j\;=\;-\sqrt{-\mbox{det}G}\;[G^{\sigma\tau}(\dot{\Phi}+A_\nu
\dot{X^{\nu}})+G^{\sigma\sigma}(\Phi^{\prime}+A_\nu\;X^{\prime\nu})].
\end{equation}
and $\rho$ by:
\begin{equation}
\rho\;=\;\sqrt{-\mbox{det}G}\;[G^{\tau\tau}(\dot{\Phi}+A_\nu\dot{X^{\nu}})+
G^{\tau\sigma}(\Phi^{\prime}+A_\nu X^{\prime\nu})]
\end{equation}
They represent the current and charge density on the world-sheet,
respectively. The continuity equation is obtained from the equation of
motion for $\Phi$:
\begin{equation}
\dot{\rho}\;=\;j^{\prime}.
\end{equation}
As a special case we now consider the de Sitter background:
\begin{equation}
ds^{2}\;=\;-(1-H^{2}r^{2})dt^{2}\;+\;\frac{dr^{2}}{1-H^{2}r^{2}}\;+\;r^{2}
d\theta^{2}\;+\;r^{2}\sin^{2}\theta d\phi^{2},
\end{equation}
supplemented by:
\begin{equation}
A\;\equiv\;A_{\mu}dx^{\mu}\;=\;0.
\end{equation}
Furthermore, we are only interested in circular strings as obtained by
the ansatz:
\begin{equation}
t=\tau,\;\;r=r(\tau),\;\;\theta=\frac{\pi}{2},
\;\;\phi=\sigma,\;\;\;\mbox{as well as:}\;\;\;\;\Phi=\Phi(\tau).
\end{equation}
It is straightforward to show that the string equations of motion take
the form for $\Phi:$
\begin{equation}
\dot{\Phi}\;=\;-\frac{\Omega}{E}(1-H^{2}r^{2})\;(\mu +\frac{\Omega^{2}}{2r^2})
\end{equation}
and for $r:$
\begin{equation}
\dot{r}^{2}\;-\;(1-H^{2}r^{2})^{2}+\frac{r^{2}}{E^{2}}\;(1-H^{2}r^{2})^{3}(\mu
+
\frac{\Omega^{2}}{2r^{2}})^{2}=0,
\end{equation}
where E and $\Omega$ are integration constants. From eqs.$(2.5)$
and $(2.6)$ it follows that:
\begin{equation}
\rho\;=\;\Omega,\;\;\;\;j\;=\;0,
\end{equation}
so that the ansatz $(2.10)$ describes a uniformly charged circular
string with charge density $\Omega$ and no current. The integration
constant E is interpreted as a constant energy density:
\begin{equation}
E\;=\;-\frac{1}{2\pi}\;\int_{0}^{2\pi}\frac{\delta{\cal
L}}{\delta\dot{t}}\;d\sigma,
\end{equation}
which, however, should not be confused with the energy density as
obtained from the spacetime energy-momentum tensor $T^{\mu\nu}.$ The remaining
physical degree of freedom of the string is the radius of the loop which
is determined by eq.$(2.12)$. For $\Omega=0$ it reduces to:
\begin{equation}
\dot{r}^{2}\;-\;(1-H^{2}r^{2})^{2}\;+\;\frac{r^{2}}{\epsilon^{2}}
(1-H^{2}r^{2})^{3}\;=\;0,
\end{equation}
where $\epsilon\equiv E/\mu,$ in agreement with the result of
Basu, Guth and Vilenkin \cite{Bas}. In what follows, however, we will only be
interested in the case of charged strings $(\Omega\neq0)$. Notice that
both eqs.$(2.15)$ and $(2.12)$ can be solved explicitly in terms of
elliptic integrals, but the general solutions will not be important here.
It is convenient to write eq.$(2.12)$ in the form:
\begin{equation}
\dot{r}^{2}\;+\;V(r)\;=\;0,
\end{equation}
where:
\begin{equation}
V(r)\;=\;(1-H^{2}r^{2})^{2}\;[\frac{1}{E^{2}}(\mu
r+\frac{\Omega^{2}}{2r})^2\;(1-H^{2}r^{2})-1],
\end{equation}
so that the classical motion takes place at the r-axis in a
$(r,\;V(r))$ diagram. The allowed loop trajectories can therefore be
extracted from the knowledge about the zeros of the potential $V(r).$ In
general we have that $V(H^{-1})=0$
and $V(r\rightarrow0)\;=\;\infty\;(\mbox{for}\;\Omega\neq0$), and the
equation $V(r)=0$ reduces to a cubic equation in the variable $\;r^{2}$.
This allows for a complete and explicit classification of the classical
motion for any values of the parameters $(H,\mu,\Omega,E)$. For our
purposes, (see Section $3)$
it is sufficient to consider the limiting case $E\rightarrow0,$ where the
potential takes the form of $\mbox{Fig}.1$, i.e. a classical circular string
with $E=0$ cannot exist inside the horizon, $r_{\mbox{hor}}=H^{-1}$.

\section{Self-Gravity and Nucleation Probabilities}
\setcounter{equation}{0}
In the preceding section we derived the potential that determines the
loop radius of a charged string in the background of de Sitter space.
Independently of the detailed form, the potential always blows up for
$r\rightarrow0$. However, the physical picture discussed in Section $2$
cannot be trusted for $r\rightarrow0$. Besides the assumption that the
thickness of the string is infinitesimal, we also have neglected the
self-gravity of the string. For a small string-loop the gravitational
effects cannot be neglected. More specifically, if the circular string
falls into its own Schwarzschild radius, it will inevitably collapse into a
black hole. The precise dynamics of a circular string including
gravitational and electromagnetic effects (self-gravity, radiation
back-reaction, energy loss due to the radiation etc.) is not known.
However, we will now argue that the effect of the self-gravity is a
"prefactor" in the form of the potential near the value of the Schwarzschild
radius of the string. Our argument goes as follows: the dynamics of a
circular string with self-gravity in a de Sitter background is somewhat
similar (can be approximated) to the dynamics of a circular string
without self-gravity in a Schwarzschild de Sitter background with an
appropriate value for the mass parameter $M$ (here considered as a free
parameter).
The metric of such a background is given by:
\begin{equation}
ds^{2}\;=\;-(1-\frac{2M}{r}-H^{2} r^{2})dt^{2}\;+\;\frac{dr^{2}}
{1-\frac{2M}{r}-H^{2} r^{2}}\;+r^{2} d\theta^{2}\;+\;r^{2}\sin^{2}\theta
d\phi^{2}.
\end{equation}
For $0\leq HM\leq1/\sqrt{27},$ the Schwarzschild de Sitter space has a de
Sitter event horizon $(r_{+})$ and a Schwarzschild horizon $(r_{-})$ which are
given by:
\begin{equation}
Hr_{+}\;=\;\frac{1}{3^{\frac{1}{3}}Z(HM)}\;+\;\frac{Z(HM)}{3^{\frac{2}{3}}}
\end{equation}
and:
\begin{equation}
Hr_{-}\;=\; \frac{1}{2}(i\sqrt{3}-1)\;\frac{1}{3^{\frac{1}{3}}Z(HM)}-
\frac{1}{2}(i \sqrt{3}+1)\;\frac{Z(HM)}{3^{\frac{2}{3}}},
\end{equation}
where:
\begin{equation}
Z(HM)\;\equiv\;[-9HM+\sqrt{3(-1+27H^{2}M^{2})}\;]^{\frac{1}{3}}.
\end{equation}
The range of values $HM>1/\sqrt{27}$ is unphysical as it
corresponds to a spacetime with a naked singularity. In what follows we
will restrict ourselves to the values
$0<HM<1/\sqrt{27}$. By a direct generalization of our
treatment of charged circular strings in de Sitter space
in the previous section, the equation of motion for the loop radius can be put
in the desired  form $\dot{r}^{2}+V(r)=0,$ where
the potential $V(r)$ is now given by:
\begin{equation}
V(r)\;=\;(1-H^{2} r^{2}-\frac{2M}{r})^{2}\;[\frac{1}{E^{2}}(\mu r
+ \frac{\Omega^{2}}{2r})^{2}(1-H^{2}r^{2}-\frac{2M}{r})-1].
\end{equation}
It follows that $V(r_{\pm})=0$, i.e. the potential from
strongly repulsive becomes locally attractive for $r\rightarrow 0.$ For
$M\ll H^{-1}$ we find the approximate expressions
\begin{equation}
Hr_{+}\;=\;1-HM\;+\;{\cal O}(H^2 M^2)
\end{equation}
and:
\begin{equation}
Hr_{-}\;=\;2HM\;+\;{\cal O}(H^2 M^2),
\end{equation}
and the potential ($3.5$) takes the form of Fig.$2$ in the limit
$E\rightarrow 0$.
In effect, self-gravity creates a finite potential energy barrier
for arbitrary values of the conserved energy $E.$ A charged
circular string has thus the possibility to tunnel quantum mechanically
through the barrier. More importantly in the limit of $E\rightarrow 0$
such a tunneling process can be interpreted as a spontaneous nucleation
of charged circular strings.  Because of charge conservation such
charged strings must be created in pairs spontaneously, and then
evolve by flying apart from each other.
The potential
barrier derived in this section is responsible for the quantum
creation process; a tunneling from nothing  into a finite radius
circular loop for each of the string loops.
It must be stressed that this nucleation process can only
be modelled through the inclusion of self-gravity of each of the
circular charged strings. This is evident by a direct comparison of
Figs.$1$ and $2$.
The probability for nucleation can now be evaluated in the WKB
approximation:
\begin{equation}
{\cal T} \;\propto\;e^{-B},
\end{equation}
where:
\begin{equation}
B\;=\;2\int_{r_{1}}^{r_{2}}\;|P_{r}|\;dr,
\end{equation}
with:
\begin{equation}
P_{r}\;=\;\int_{0}^{2\pi}\;\;\frac{\delta{\cal L}}{\delta\dot{r}}
d\sigma.
\end{equation}
Here $r_{1}$ and $r_{2}$ are the turning points and ${\cal L}$ is the
Lagrangian that describes the charged string. As we have already pointed
out, the exact dynamics of the circular strings is not known.
The Schwarzschild de Sitter model that led to the potential $(3.5)$ is
therefore to be taken as an approximation only.
A similar approximation, which we will use in what follows, is based on
the de Sitter model that naturally led to the potential $(2.17),$
but with a physical cut-off (due to the self-gravity)
that pulls down the rising repulsive
potential at some value of the radius $r=\delta\ll H^{-1}$. In the following
we consider $\delta$ as a parameter to be measured experimentally. In
eqs.$(3.8)-(3.10)$ we then get:
\begin{equation}
r_{1}=\delta,\;\;\;\;r_{2}=H^{-1},\;\;\;\;|P_{r}|\;=\;2\pi\frac{\mu
r+\Omega^{2}/2r} {\sqrt{1-H^2 r^2}},
\end{equation}
in the limit $E\rightarrow 0$. It follows that :
\begin{equation}
B\;=\;\frac{4\pi\mu}{H^{2}}\sqrt{1-H^{2}\delta^{2}} +
2\pi\Omega^{2}\ln\frac{1+\sqrt{1-H^2\delta^{2}}}{H\delta}.
\end{equation}
For $H\delta\ll 1$ we find that:
\begin{equation}
B\;=\;-2\pi\Omega^{2}
\ln(H\delta)+2\pi(\frac{2\mu}{H^{2}}+\Omega^{2}\ln{2})
\;+\;{\cal O}(H^2\delta^{2}),
\end{equation}
i.e. :
\begin{equation}
{\cal T}
\;\approx\;(H\delta)^{2\pi\Omega^{2}}\;\;\mbox{exp}[-\frac{4\pi\mu}{H^2}(1+
\frac{\Omega^{2}H^{2}}{2\mu}\ln2)].
\end{equation}
For $\delta=\Omega=0$, we observe that the nucleation amplitude reduces
to the one for the uncharged string \cite{Bas} (as it should). In
the more general case
under consideration here it is convenient to introduce the dimensionless
variables:
\begin{equation}
x\;\equiv\;H\delta\ll 1,\;\;\;\; y\;\equiv\;H^{2}\Omega^{2}/\mu\;\geq\;0
\end{equation}
and to plot the function:
\begin{equation}
F(x,y)\;=\;\frac{H^2}{2\pi\mu}\ln\frac{{\cal T}_{BGV}}{{\cal T}}\;=\;
2(-1+\sqrt{1-x^2})\;+\;y\ln\frac{1+\sqrt{1-x^2}}{x},
\end{equation}
where  ${\cal T}_{BGV}=e^{-4\pi\mu/H^2}\;$
(see Fig.$3$.)\\
                \\
\\
Some comments are now in order:\\
                                  \\
1.The original Basu-Guth-Vilenkin (BGV) result \cite{Bas}
corresponds to the point $x=y=0.$                   \\
                                                                         \\
2. For a zero charge string $(y=0)$ the nucleation probability is
not sensitive to the physical cut-off $\delta,$ i.e.
the BGV result is essentially
independent of the self-gravity of the string.\\
\\
3. For a charged string $(y\neq 0)$ the nucleation probability is
extremely sensitive to the cut-off. More specifically it
goes to zero as a power of $x$ for $x\rightarrow 0$.\\
\\
4. By now combining $2$ and $3,$ $F(x,y)$ is singular along the
$y$-axis, where ${\cal T}=0$, except for $y=0$ where
${\cal T}={\cal T}_{BGV}$. On the other hand,
$F(x,y)$ is regular on the $x$-axis for $0\leq x<1$ (where $x$ is defined).\\
\\
With the above  comments our discussion of the charged string nucleation
probability is complete. We will finish by making a few observations about
the classical evolution of circular charged strings after nucleation. As
they nucleate at horizon size, their evolution is most conveniently
described in the comoving time :
\begin{equation}
T\;=\;\frac{\ln|1-H^2r^2|}{2H}\;+\;t.
\end{equation}
Notice that $r$ equals the physical radius as seen in spatially flat
Robertson-Walker coordinates.
After a little algebra we find from eq.$(2.12)$ when $E\rightarrow 0$:
\begin{equation}
\frac{dr}{dT}\;=\;-\frac{1-H^2 r^2}{H r},
\end{equation}
which is solved by:
\begin{equation}
H\;r(T)\;=\;\sqrt{1+e^{2H(T-T_{\circ})}},
\end{equation}
thus the string expands with the same rate as that of the universe (for
$H(T-T_{\circ})\gg 1)$. Interestingly, this result is independent of
both the string charge $\Omega$ and the string tension $\mu.$ In fact, the
physics here is identical with the evolution of the uncharged circular
string as originally was discussed by Basu, Guth and Vilenkin \cite{Bas}. The
reason is that the different parts of the string loop once outside the
horizon are not causally connected. The system simply follows the
expansion of the universe.
\section{Conclusion}
In the present work we investigated the dynamics of circular charged
strings in a de Sitter spacetime background. We derived the radial
equations of motion and identified the repulsive classical potential
energy due to the rapid expansion of the background. We included the
effect of self-gravity for small radius of the loop by deriving
the corresponding equations of motion in a Schwarzschild de Sitter
background. We demonstrated the appearance of a classical potential energy
barrier and computed the probability for a quantum tunneling
of a single uniformly charged loop with zero energy.

As explained in Section 3, we
interpret such a spontaneous nucleation effect as the quantum
creation of a pair of oppositely charged circular string loops at the
cosmic horizon and their subsequent parting off.
\section{Acknowledgements}
The research of M.A. was partially supported by Danmarks
Grund-Forsknings Fond through its support for the establishment of the
Theoretical Astrophysics Center.

\newpage
\begin{centerline}
{\bf Figure Captions}
\end{centerline}
\vskip 24pt
\hspace*{-6mm}Fig.1. The potential $(2.17)$ determining the loop-radius of a
charged string in de Sitter space. Shown is the case where $E=0.$ The string
motion is confined to the $r$-axis, i.e. the string cannot exist inside the
horizon.
\vskip 12pt
\hspace*{-6mm}Fig.2. The potential $(3.5)$ determining the loop-radius of a
charged string in Schwarzschild de Sitter space. As in Fig.$1$ we consider
only the case where $E=0.$
\vskip 12pt
\hspace*{-6mm}Fig.3. The function $F(x,y)$ (eq.$(3.16)$) determining the
nucleation probability of charged strings in de Sitter space when
including the self-gravity. Notice that
$F=0$ corresponds to ${\cal T}={\cal T}_{BGV}$
and $F=\infty$ corresponds to ${\cal T}=0.$

\newpage
\begin{thebibliography}{11}
\bibitem{Linde} A. Linde, `Particle Physics and Inflationary Cosmology'
 (Harwood, Chur, 1990).
\bibitem{Rocky} E.W. Kolb and M.S. Turner, `The Early Universe'
(Addison-Wesley 1989).
\bibitem{RB} R.H. Brandenberger in `Lectures in Modern Cosmology and
Structure Formation', proc. of the 7th Swieca Summer School (World
Scientific, Singapore, 1993).
\bibitem{Bas} R. Basu, A.H. Guth and A. Vilenkin, Phys.Rev.D44 (1991) 340.
\bibitem{JG} J. Garriga and A. Vilenkin, Phys.Rev.D47 (1993) 3265.
\bibitem{SH} S.W. Hawking, Phys.Lett.B246 (1990) 36.
\bibitem{Wit}E. Witten, Nucl.Phys.B291 (1985) 557.
\bibitem{Jer}J.P. Ostriker, C. Thomson and E. Witten,
             Phys.Lett.B180 (1986) 231.
\bibitem{Car}B. Carter, Phys.Lett.B224 (1989) 61.
\bibitem{Cop}E. Copeland, M. Hindmarsh and N. Turok, Phys.Rev.Lett.58 (1987)
             1910.
\bibitem{Bal}A.P. Balachandran, B.S. Skagerstam and A. Stern, Phys.Rev.D20
             (1979) 439.
\bibitem{Dav}A. Davidson and K.C. Wali, Nucl.Phys.B348 (1991) 581.
\bibitem{Vil}A. Vilenkin and T. Vachaspati, Phys.Rev.Lett.58 (1987) 1041.
\bibitem{Spe}D.N. Spergel, T. Piran and J. Goodman, Nucl.Phys.B291 (1987) 847.
\bibitem{All}A.L. Larsen, Class.Quant.Grav.10 (1993) 1541.
\bibitem{JK}A. Davidson, N.K. Nielsen and Y. Verbin,
            Nucl.Phys.B412 (1994) 391.
\end{thebibliography}
\end{document}